# An energy-based macroeconomic model validated by global historical series since 1820

**Authors:** Hervé Bercegol[1,3] *, Henri Benisty[2,3].

**Affiliations:**

[1] SPEC, CEA, CNRS, Université Paris-Saclay, CEA Saclay, 91191 Gif-sur-Yvette, France.

[2] Laboratoire Charles Fabry, Institut d'Optique Graduate School, CNRS, Université Paris Saclay, 2 Avenue Augustin Fresnel, 91127 Palaiseau Cedex, France.

[3] Laboratoire Interdisciplinaire des Énergies de Demain (LIED), CNRS UMR 8236, Université Paris Diderot, 5 Rue Thomas Mann, 75013 Paris, France.

*Correspondence to: herve.bercegol@cea.fr

**Abstract:** Global historical series spanning the last two centuries recently became available for primary energy consumption (PEC) and Gross Domestic Product (GDP). Based on a thorough analysis of the data, we propose a new, simple macroeconomic model whereby physical power is fueling economic power. From 1820 to 1920, the linearity between global PEC and world GDP justifies basic equations where, originally, PEC incorporates unskilled human labor that consumes and converts energy from food. In a consistent model, both physical capital and human capital are fed by PEC and represent a form of stored energy. In the following century, from 1920 to 2016, GDP grows quicker than PEC. Periods of quasi-linearity of the two variables are separated by distinct jumps, which can be interpreted as radical technology shifts. The GDP to PEC ratio accumulates game-changing innovation, at an average growth rate proportional to PEC. These results seed alternative strategies for modeling and for political management of the climate crisis and the energy transition.

## 1. Introduction

Energy resources are essential to provide wealth and quality of life to human societies[1,2]. Any economic process consumes energy, i.e. turns energy from a valuable, low-entropy form into a high-entropy, waste form[1]. Nowadays, fossil carbon sources still provide about 85% of primary energy consumption (PEC), remaining the principal driver of climate change through carbon dioxide ($CO_2$) emissions, as identified since the 1970s[3,4]. New renewable energy technologies (photovoltaic panels, wind turbines), deployed for decades, recently became competitive with fossil fuels in several sectors: electricity[5], urban mobility, thermal management of well-insulated buildings, to name a few. Additional technologies in other services, notably chemical production from non-fossil resources like atmospheric $CO_2$,[6] justify strong programs to spread low-emission energies[7]. Despite propositions for all-out deployment of fossil-free sources[8,9], these technologies, however crucial to our future wealth and wellbeing, still make a modest piece of the energy pie[10]. Currently, the macroeconomic field is lacking a universally accepted model that would give energy its fair share as a systemic input. Scenarios such as those reviewed by the Intergovernmental Panel on Climate Change[11] are based on "integrated assessment models" where energy is a mere sector of the economy, no larger than its nominal cost share (5 to 10%). Common economic textbooks[12,13,14] generally script capital and labor as the two key factors of production, with a correction factor (residual in the neoclassical Solow model[12-14]) accounting for the positive effect of knowledge development. Ecological and biophysical economists have long criticized[15,16] the negligible role given to energy as unrealistic, leveraging[17,18] the evidence from national accounting (USA, Japan, Germany, etc.) that the energy consumption share is much more important in production functions, about as important as physical capital. Knowing this, macroeconomic models for the energy transition should cease ignoring the systemic role of energy embodied in capital and labor, as repeatedly demonstrated[19,20,21,22].

In this paper, we present a simple, energy-based macroeconomic model to study the world economy, as an essential step to unravel the **economy-environment nexus**. Inspired by ecological economics[1,15-22] and neoclassical macroeconomics[12-14], this model aims at reconciling both schools. Since we are aiming for a deep underlying relation between GDP and PEC, our approach fundamentally differs from a vast literature[20,23] that looks for signs of causality from PEC to GDP or vice versa. In particular Granger causality, obtained through statistical treatment, often proved inconclusive[24]. Moreover, papers like ref. 23 generally focus on relations at national or nation-block level, studied on rather short series of a few decades, contrary to our global assessment in the long-run.

## 2. A new assessment of global production and energy data

The time scale of the aggregated world economy involves decades: thus, its understanding can only rest on long-run observations, spanning centuries, as done for the climate system itself. Prompted by the current climate crisis and the emerging Anthropocene concept, developing and probing a new macroeconomic model makes sense from the recent availability of the thorough, painstaking data collections and authoritative analyses of energy consumption[25,26,27]. The same holds for the estimates of global scale economic production in the long-run, pioneered by Angus Maddison[28] and his successors[29]. appendix A discuss the various data sources. GDP data show an overall homogeneity while PEC sources suffer from some discrepancies: data from ref. 25 are used in this paper, preferred to other published series[30,31] for reasons detailed in appendix A.

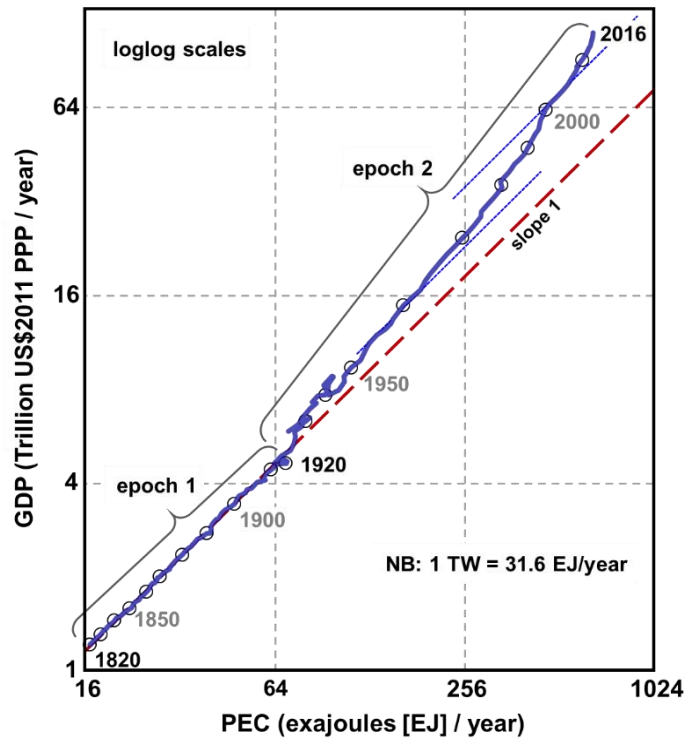

**Fig. 1**. **Log-log plot of Global annual GDP $Y$ vs. Annual Primary Energy Consumption (PEC) $E$**. GDP is in US$-2011 at parity of purchasing power (see appendix A). The time scale is marked as empty disks at decades 1820, 1830 up to 2010. Energy aggregates all types of PEC, by humans as food, by draft animals as fodder, by diverse fireplaces or machines as fuels of any sort or wind or sun streams. In epoch 1, from 1820 to 1920, both $Y$ and $E$ vary by a factor ~4; a dashed red line extrapolates their linear relation, unit slope in this log-log plot, through two successive factors 4. During the following century (epoch 2, 1920-2016), $Y$ grows quicker than $E$, while still following $E$ quasi linearly (unit slope e.g. marked by blue dotted lines) for periods of a few years. See §3.1 and §3.2 for explanation of both features. (1 Exajoule [EJ] is $10^{18}$ J; a Trillion $ equals $10^{12}$ $; a Terawatt [TW] is $10^{12}$ W).

Figure 1 shows the sustained growth pattern of the world annual PEC $E$ and GDP $Y$. During the last 200 years, GDP grew by a factor of about 90, while PEC was multiplied by 40. Both GDP and PEC have grown exponentially with time, at variable rates in different periods. Choosing period durations of a few decades (between 20 to 36 years, see table 1), the average annual growth never went under ~1 %.

**Table 1**. Relative growth rates of annual PEC $E$, global population $L$ and world annual GDP $Y$.

| periods | Average Annual Growth Rate | | | epochs (see Fig. 1) | Relative growth |
|---|---|---|---|---|---|
| | $E =$ Global PEC | $L =$ Global Population | $Y =$ Global GDP | | |
| 1820-1850 | 1.0% | 0.5% | 0.9% | epoch 1 | $\dfrac{E_{1920}}{E_{1820}} = 4.2 \quad \dfrac{L_{1920}}{L_{1820}} = 1.8 \quad \dfrac{Y_{1920}}{Y_{1820}} = 3.8$ |
| 1850-1870 | 1.1% | 0.3% | 1.2% | | |
| 1870-1900 | 1.8% | 0.7% | 1.8% | | |
| 1900-1920 | 1.9% | 0.8% | 1.5% | | |
| 1920-1950 | 1.6% | 1.0% | 2.4% | epoch 2 | $\dfrac{E_{2016}}{E_{1920}} = 9.5 \quad \dfrac{L_{2016}}{L_{1920}} = 4.0 \quad \dfrac{Y_{2016}}{Y_{1920}} = 24$ |
| 1950-1980 | 3.7% | 1.9% | 4.5% | | |
| 1980-2016 | 1.9% | 1.4% | 3.1% | | |

The main point emerging from the careful treatment of GDP and PEC data (see details in appendix A) is the overall nearly linear trend on a log-log scale, over about 2 orders of magnitude, with a near-unity exponent[25] in epoch 1 (1820-1920, Fig. 1 and Table 1). Note that in this epoch there are already boom and bust, but they only scatter the underlying time along the line. The trend differs in epoch 2 (1920-2016), consisting of a piecewise patch of segments of similar unit slope, interspersed with jumps. We now lay out a modeling frame that accounts for the distinct features of both epochs.

## 3. Results and model significance

*3.1 Epoch 1 (1820-1920): reassessment of production factors*

The regular, linear relation observed in epoch 1 is interpreted using rather standard macroeconomic growth modeling[12-14]. $Y$ is described by an aggregate production function $Y = f(K, H, L)$ which a priori depends on physical capital $K$, human capital[32] $H$ and labor $L$, $L$ being usually identified with the population number. As introduced in appendix B, and generally adopted[12-14], $H$ capitalizes the knowledge and skills of workers, while $K$ represents the tangible manufactured goods and infrastructure that support production. In ref. 12-14, the standard treatment considers an exponential growth of $L$ which induces a similar proportional growth of $Y$. However, as evidenced in Fig. 1 and table 1, the proportionality lies here between $Y$ and $E$, not $L$. A very simple, although bold operation – embedding $L$ into $E$ – provides a strong link between standard modeling and global data. This new viewpoint springs from the understanding that labor – unskilled human work fueled by food – assimilates to any of the other forms of energy consumption[22] as the physics term "work" points out. Indeed human labor has continuously been enhanced by new energy sources and converters throughout the Homo genus records. Enhancing $L$ by energy consumption per capita $\varepsilon = \frac{E}{L}$, we consider a production function of the "Cobb-Douglas"[12-14,18,22] type:

$$\frac{Y}{Y_{1820}} = A \left(\frac{K}{K_{1820}}\right)^\alpha \left(\frac{H}{H_{1820}}\right)^\beta \left(\frac{\varepsilon L}{\varepsilon_{1820} L_{1820}}\right)^\gamma = A \left(\frac{K}{K_{1820}}\right)^\alpha \left(\frac{H}{H_{1820}}\right)^\beta \left(\frac{E}{E_{1820}}\right)^\gamma \quad (1)$$

where factors are normalized to their value in 1820. $\alpha$, $\beta$ and $\gamma$ are the elasticities of production of $K$, $H$ and $E$; inequalities $0 < \alpha < 1$, $0 < \beta < 1$ and $0 < \gamma < 1$ are standard hypotheses of diminishing returns; the condition $\alpha + \beta + \gamma = 1$ provides for constant return-to-scale with respect to the three factors. Interpreting Eq. (1) according to standard macroeconomic theory[12-14], $\gamma$ is the global cost share of energy needs (including basic needs of workers and families, mainly food) while $\beta$ (resp. $\alpha$) is the part of production rewarding human (resp. physical) capital.

The *residual* $A$,[12-14] also called *total factor productivity*, takes care of any discrepancy between the products of factors and $Y$. $A$ equals unity in 1820 – by definition – and we keep it so for epoch 1. As we will discuss later, $A$ evolves in epoch 2.

$K$ and $H$ are *stock* variables whereas $E$ stands for a *flow*, the consumption of energy – actually a consumed *power* expressed in power units, Exajoule/year (EJ/year) or TW (1 TW=31.6 EJ/year). At a given time, $K$ and $H$ "store" the results of history, actually through the integral of PEC $E$, but also through $K$ and $H$, previously invested into machine or plant building and worker formation. The dynamics in this system is set by the evolution equations of the accumulated capital $K$ and $H$. $K$ (resp. $H$) grows with time t as a share $s_K Y$ (resp. $s_H Y$) of the production, saved and invested, while a part $\delta_K K$ (resp. $\delta_H H$) is lost by depreciation.

$$\frac{dK}{dt} = s_K Y - \delta_K K \quad \& \quad \frac{dH}{dt} = s_H Y - \delta_H H \quad (2)$$

with $s_K$ and $s_H$ (dimensionless, between 0 and 1) and $\delta_K$ and $\delta_H$ (inverse times) the savings and depreciation rates. Equations 1 and 2, with $A = 1$, constitute a model[12-14,32] worth further studies (cf. appendix B). However, aware of the scarcity of data for $K$ and $H$ in epoch 1, we stick to simplicity, in a system where normalized production $\mathcal{Y}$ is a function of only two factors,

$$\frac{Y}{Y_{1820}} = \mathcal{Y} = \mathcal{J}^\alpha \mathcal{E}^{1-\alpha} \quad (3)$$

$\mathcal{E}$ is the normalized PEC, extrinsically defined, and $\mathcal{J}$ is an aggregated capital factor accumulating according to:

$$\frac{d\mathcal{J}}{dt} = s\mathcal{Y} - \delta \mathcal{J} \qquad (4)$$

$\mathcal{J}$ is a combination of $K$ and $H$, and can stand for both of them, or for any factor that requires PEC to accumulate and build up. Equations 3 and 4 constitutes a simple system thoroughly studied in macroeconomics[12-14]. When $\mathcal{E}$ grows exponentially as $\exp[gt]$ (see exact solution in appendix B), they give rise to a steady-state *balanced growth path*[12-14] where both $\mathcal{Y}$ and the aggregated factor $\mathcal{J}$ grow exponentially at the same rate as $\mathcal{E}$, maintaining a constant ratio $\frac{\mathcal{J}}{\mathcal{Y}} = \frac{s}{g+\delta}$, whatever the value of $\alpha$ in Eq. (3). In a nutshell, as $\mathcal{J}$ capitalizes production, it boils down to storage of energy. However, coming back to the question of causality evoked in the introduction, it would be erroneous to conclude that there is consequently a unidirectional causality from PEC to GDP. Conservatively, one had better describe it as a bidirectional causality or even a "chicken and egg" situation. Our conviction – coming from the physics of open complex systems – is that the intricacy of the global production function and its factors rather requires a systemic description that goes beyond causality, not precluding the relevance of causality studies in appropriate subsets[23-24,33]. In this balanced growth path, PEC growth induces GDP growth from Eq. (1); simultaneously, a rise in GDP causes capital to rise quicker (Eq. (2)), and a larger capital implies a bigger PEC to feed it. This last assertion may stop being true when new, radically more efficient technologies replace old ones, a situation that is relevant to epoch 2, discussed in §*3.2*.

Epoch 1 in Fig. 1 evidences such a balanced growth phenomenon, with the iconic decades of the coal-iron-railway "development block" thoroughly described by Kander et al.[26]: coal transforms iron, iron makes rail and locomotives, railways transport coal, in a growing spiral fueled by coal burning. To grasp the specific role of $K$ and $H$ growth, it is interesting to take the log-differential of Eq. (1) which, since $Y \propto E$ here, leads to $\frac{\Delta K}{K} \simeq \frac{\Delta Y}{Y} + \frac{\beta}{\alpha}\left(\frac{\Delta Y}{Y} - \frac{\Delta H}{H}\right)$.

Since Piketty[34] reports that global capital $K$ grows slightly quicker than $Y$ in epoch 1 (see ref. 34, p. 196), we infer that $H$ should have experienced a slower growth than $Y$ in those periods, which is a testable assumption. It should translate into a variation of saving and/or depreciation parameters.

*3.2 Epoch 2 (1920-2016): how energy feeds Schumpeterian innovation*

In epoch 2, a different regime of economic growth settles. The 1920-1950 period, with the roaring twenties, Great Depression and WWII appears in Fig. 1 as a transient towards a new energy/GDP relation. After exploring some alternatives, we deduce from the structure of Fig. 1 and the physical justification of the linearity of $Y$ and $E$ a closed-form relation applying to the whole timescale, which is a boldly simplified version of Eq. (1):

$$\frac{Y}{Y_{1820}} = A \frac{E}{E_{1820}} \quad \text{or} \quad \mathcal{Y} = A \mathcal{E}, \tag{5}$$

The time-dependent residual $A$ carries the whole flow of innovation spurred by the use of energy – tapping new resources, efficiency gains in converting PEC to *useful work*[21], inventing new conversion modes, machines, materials or distribution schemes – but also any revolutionary progress of universal knowledge (electricity, internal combustion engine, air travel). $A = \mathcal{Y}/\mathcal{E}$ can be considered as the *normalized productivity of energy*[25]. In Eq. 5, where the linearity between $Y$ and $E$ hides a relation like Eq. (1), $H$ and $K$ are implicitly present, merged into $E$ along the exponential, balanced growth path. A more complex modeling is explored in SM, keeping capital $\mathcal{J}$, which turns out to produce similar results.

There is a vast macroeconomic literature on such residuals in Cobb-Douglas-type production functions. We favor an endogenous definition of $A$,[35] and look for a simple factor dependency. Imitating similar considerations on the long run human history[36], and guided by the linear $Y$ to $E$ relation at work during epoch 1, we merely conjecture proportionality between $E$ and the average growth rate of $A$

$$\frac{dLog[A]}{dt} = \chi \mathcal{E}, \tag{6}$$

which could also be written $\frac{dA}{dt} = \chi \mathcal{E} A = \chi \mathcal{Y}$. Since $A$ contains all irreversible progress eventually shared by the whole humanity, so-called *nonrival goods*, it seems logical that any effort (measured in energy unit) anywhere on the planet has the same average, long-term effect on $A$ growth.

From Eq. (6), we derive $A[t] = \exp[\int_{1820}^{t} \chi \mathcal{E} \, dt]$ since $A[1820]=1$. If one remembers that here $\mathcal{E}$ contains unskilled labor $L$, then Eq. (6) resembles Kremer's bold assumption[36], i.e. that $\frac{dA}{dt}$ was roughly proportional to $LA$. Figure 2 shows that, between $A = \mathcal{Y}/\mathcal{E}$ and the time integral of PEC ($\int_{1820}^{t} \mathcal{E} \, dt$), an exponential relation holds from 1930 on, with

$\chi \approx 0.00046 \, \text{year}^{-1}$ (this value hinges on the 1820 normalization of PEC in $\mathcal{E}$), hence the nonlinear skew only in epoch 2. We insist once again that the time integral yields a bona fide energy, since the PEC $E$ is an energy flow, a physical power indeed. During the 1920s and 1930s, $\mathcal{Y}/\mathcal{E}$ jumps from the epoch 1 conditions to the new epoch 2 regime, with the full-fledged range of modern techniques substituting older ones. Projecting the exponential relation between $A$ and $\int_{1820}^{t} \mathcal{E}\, dt$ to the past, we see that the new regime integrates the effect of energy consumption on invention from the beginning of 19th century. Innovation in energy efficiency, and introduction of new machines etc. existed in epoch 1. Its first 50 years, described as the age of steam and railways, would rightly be characterized by a single trend in energy usage, with the deployment of steel and electrical power in the 1870s and 1880s qualifying as a new stage of industrial revolution[2,21,37].

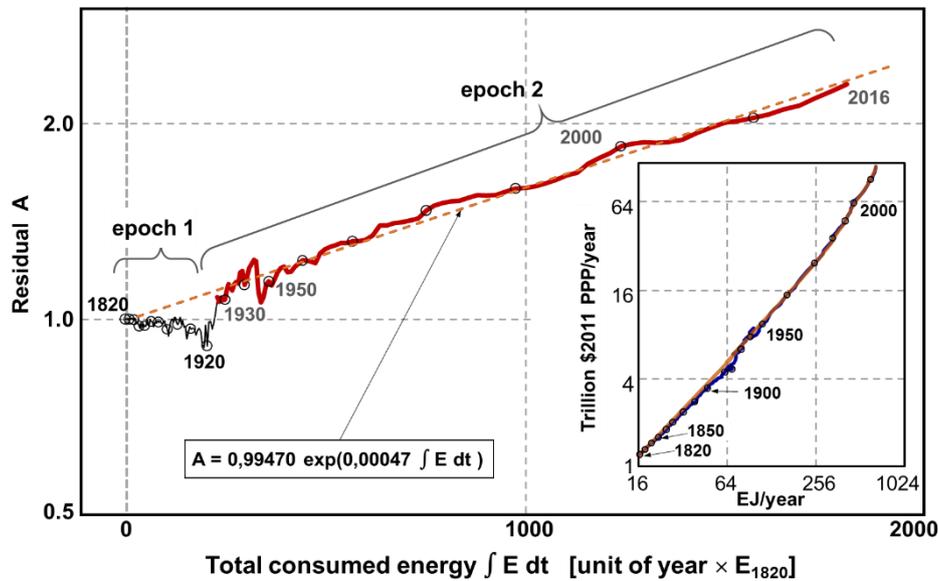

**Fig. 2. Semilog plot of the residual** $A$, cf. Eq. (5), vs. the time integral of PEC, normalized to $E[1820]$. An exponential relation holds between 1930 and the present. From the beginning of 20th century, $A$ jumps from its previous value of 1 in the 19th century to the new regime. The exponential fit is done between 1926 and 2016. Decennial data are marked with black circles. The inset shows the data of Fig. 1 fitted with Eq. (7).

However, these new devices and industrial organization did not change the overall energy efficiency of the global economic system, maybe because a higher amount of energy had to be invested in the machines. Actually, table 1 shows that the growth rate of PEC grew steadily throughout epoch 1, evidencing a global rebound effect prompting the addition of energy thirsty processes. We would interpret this in saying that radical innovation was hindered in 1820-1920 and its path suddenly tipped in the 1920s, spurred in all domains from e.g., air transport to agriculture[38] and to energy distribution. Several successive jumps occurred during epoch 2. In the second half of the 20th century, $A$ grows by steps of a few percent as shown by the peaks of annual $\frac{\Delta A}{A}$ in Fig. 3. We account for these as radical, "Schumpeterian" technology shifts[2,21], bringing a sudden increase in energy efficiency by the abrupt replacement of less efficient schemes or devices (even though singling out specific items, however iconic, is irrelevant, as triggering effects arise from their intertwined use). However, in the last 10 years, the growth of $A$ tends to be more regular, maybe revealing a more decentralized innovation system. Nevertheless, during the whole 90 last years to date of epoch 2, the quantitative dependency of economic growth on energy does not experience any change, thus following Eq. (5) recapped as

$$\frac{Y}{Y_{1820}} = \exp\left[\int_{1820}^{t} \chi \frac{E}{E_{1820}} dt\right] \frac{E}{E_{1820}} \qquad (7)$$

Visually, we refer to the inset of Fig. 2 where we fit with Eq. (7) the data from Fig. 1.

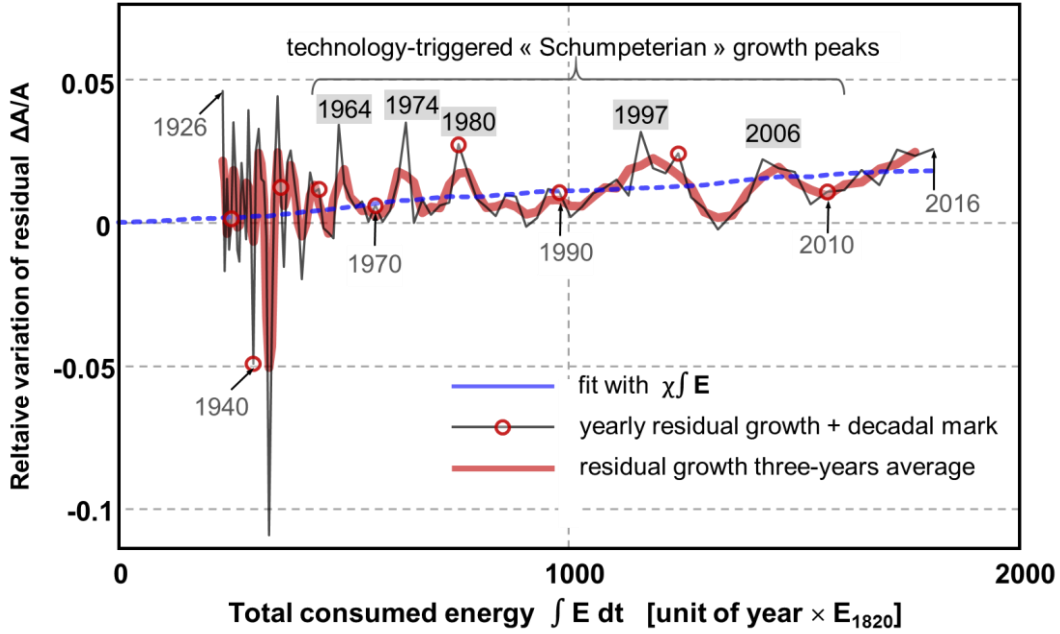

**Fig. 3. Annual relative growth rate of** $A$, residual carrying the effect of innovation on GDP, plotted vs the time integral of consumed energy, in epoch 2 (1926-2016). Jumps identified in Fig. 1 are evidenced here as peaks of $\frac{\Delta A}{A}$. Five of these peaks are marked with the years of highest innovation growth, in 1964, 1974, 1980, 1997 and 2006. Thin line: year to year relative growth rate of $A$; thick line: 3-year average. Dashed line: linear relation in $\mathcal{E}$ according to Eq. (7).

Let us now comment Fig. 3 in more details with its peaks in the residual's growth $\frac{\Delta A}{A}$, plotted for epoch 2 as well as its three-years running average. Apart from the 1940 and 1945 expected dips (obvious for WWII, whereby all factors including knowledge concur to physically destroy belligerents capital, e.g., energy networks, and labor), the average is around 1% - see the dashed line $\chi \mathcal{E}$ in Fig. 3 - with a baseline around 0.5%. But the Schumpeterian peaks appear as well, spurring typical booms of 1.5-2% extra growth in the moving average. We shall argue elsewhere about the collective "swarming" nature of these booms, as no single sector of economy can support it alone, mainly because epoch 2 economic success spawns so many new objects[39].

*3.3 Prospective trajectories for the energy transition*

Finally, a deceivingly simple exercise consists in prolonging the long-time trend with Eq. (7), ignoring the jumps. We acknowledge that the linear $Y$ to $E$ relation masks a complex function involving all capitalizing factors – see appendix B–. However we prefer a simple algebra that better emphasizes our scope, through the derivative form of Eq. (7) $\frac{\Delta Y}{Y} = \frac{\Delta E}{E} + \chi \mathcal{E}$. Then, a *business as usual* scenario, shown as a black curve in Fig. 4, merely extrapolates the ~ 1.9% PEC growth observed during the last forty years.

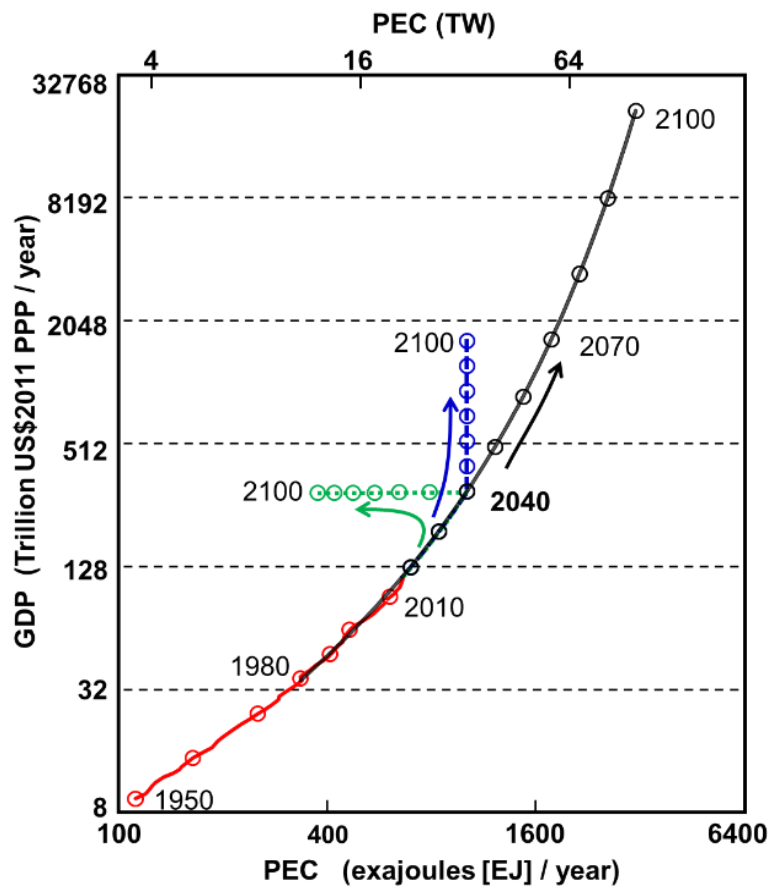

**Fig. 4. Prospective in the $E$, $Y$ plane (loglog scales), according to Eq. (7).** The black line is business-as-usual with PEC growing exponentially until 2100 at the same average rate as observed (red line) in 1980-2016. The two scenarios discussed in the text are illustrated, assuming a sharp turn, arbitrarily from 2040 on, with either constant PEC (dashed blue) or constant GDP (dotted green).

Fig. 4 also shows two other scenarios, better addressing the GDP-environment nexus. Firstly, allowing for a continued growth in PEC $E$ during a short period (until 2040 in Fig. 4), we could envision a time when the GDP "overgrowth" $\chi \mathcal{E}$ is large enough to sever any further PEC growth. This scenario would eventually usher into a new circular economy era, whereby GDP growth solely results from game-changing innovations (vertical blue segment till 2100). Secondly, Eq. (7) could also introduce a degrowth trajectory, whereby $E$ decreases while $Y$ is maintained constant from a future time $t_0$ (2040 again in Fig. 4). Such a trend in Eq. (7) would induce an algebraic decay of PEC $E = \left( \dfrac{1}{E_0} + \dfrac{\chi(t-t_0)}{E_{1820}} \right)^{-1}$, innovation gains compensating for capital and PEC degrowth. These stylized scenarios are meant as draft tools to elaborate policies at a regional or global level, adding to recent work on a possible decoupling of GDP from resources consumption[40,41].

### 4. Conclusions

The tale of epoch 1 is an energy-geared capital accumulation ($K$ and $H$) while epoch 2 sees game-changing innovation enhancing the impact of PEC. This double role of energy consumption, at the time of disquieting growth externalities in climate, biodiversity, etc., seems to us as a key driver worth considering to address the economy-environment nexus. Thanks to its dual inspiration from ecological economics[1,15-22] and neoclassical macroeconomics[12-14,32,35,36], substantiated by historical accounts[2,25-28,30,31,34,37], our model offers an opportunity for reconciling both schools. It shall also foster operational tools to manage the planet's human future, part of a conversation on a desirable energy transition trajectory. In addition to taking into account the energy-based mechanisms suggested by this study, an improved model might add money and financial flows as effective levers for capital and knowledge in essential global shifts[37].

**Acknowledgments:**

We thank Berengère Dubrulle, Carina Faber, François Graner, Roland Lehoucq, Sawako Nakamae for useful suggestions on the manuscript, and Robert Ayres for helpful comments. Both authors acknowledge the seminal role of Luc Valentin in prompting their collaboration.

**Data and materials availability:** all data are publicly available as detailed in appendix A.

**Appendices**

*Appendix A: Choice and description of historical series    pp. 13-22*
    *Text:*             *pp. 13-16*
    *Figures A1 to A6:*  *pp. 17-22*

*Appendix B: Modeling                                      pp. 23-30*
    *Text:*             *pp. 23-28*
    *Figures B1 to B2:*  *pp. 29-30*

### *A. Choice and description of historical series*

In appendix A, we present the various sources of historical series for primary energy consumption (PEC) and GDP, and justify the choice that we made among them.

#### *A.1. Global primary energy consumption (PEC)*

We have been using Paolo Malanima's recently published (Supplementary materials and downloadable database in ref.25) database, which compiles global primary energy consumption (PEC) of all energy types and sources from 1820 to 2016. Previously, the 19$^{th}$ century period was described only by scarce or partial data, apart from a compilation by Victor Court[30] that was also considered for our work (See appendix of ref. 30 p. 216-224 for the database), and will be compared to ref. 25 and its supplementary materials in the following. Malanima's main paper[25] rightly focusses onto PEC up to 1913, putting them in a geographical and historical perspective. The epoch 2 period (1920-2016) was much better known than epoch 1 (1820-1920), which justifies concentrating on the latter. However, Malanima's database provide historical series for the whole period (1820-2016) for the 8 comprehensive types of primary energy sources:

1. food for humans
2. fuelwood (which includes all solid biofuels, crop residues and derivates from wood, e.g. charcoal)
3. fodder for working animals
4. coal (all solid fossil fuels, also including peat)
5. oil
6. natural gas
7. electricity (from water, wind, geo, Sun and other renewables)
8. nuclear

Since these data relate to primary energy, electricity from coal (resp. nuclear or other fossil sources) will appear as spent coal (resp. nuclear …) and be accounted as coal's thermal energy content. Food, fuelwood and fodder (1 to 3) are traditional energy sources; we call items 4 to 8 "modern energy sources".

The energy crisis in the seventies and concerns about global warming prompted the survey of fossil fuel consumption in the long run, as well as new fossil free alternatives. Etemad and Luciani published[42] in 1991 an estimate of modern energy consumption since 1800 that has been used as a reference/basis by all authors afterwards. Fuelwood – solid biofuels – were discarded by most economic surveys up to the 2010s (being either invisible in trade or plainly ignored by surveys), although wood was a key energy resource. This is especially true during the first century of the industrial revolution, when wood fueled e.g. a major part of the 19$^{th}$ century development of the USA[43] (cf. data on p.341), a would-be superpower. Smil[31] added fuelwood (biomass) consumption since 1800, based on the work of Fernandes et al[44] who assessed the use of biofuels from 1850 to 2000. Ref. 44 is the original source of most recent estimate of fuelwood/biofuel energy use. However, Fernandes et al provide a mass of consumed biofuels – wood, wood derivatives and crop residues - by macro-area (different geographical divisions than Malanima's). A conversion coefficient is necessary to convert those masses into their primary energy content. Malanima used 12.5 MJ/kg (3000 kcal/kg) for wood, which corresponds to wood with about 33% humidity. We come back to this a little later.

Up to the collective work of Kander, Malanima and Warde[24], most published estimates of energy consumption did not include food nor fodder. The inclusion of food – energy directly consumed by human workers – is crucial for a physical understanding of the role of energy in the economy, as previously claimed in §3. The same is true for fodder – energy necessary for draft animals to fuel their [horse] power – and fuelwood, which was the main energy source before the fossil fuels era. When draft animals replace humans, or when coal-fired locomotives replace horse traction, the economic impact must be assessed together with the increase of consumption that these changes induce.

Before the inclusion of traditional energy sources into the aggregate consumption, the productivity of energy ($A = \mathcal{Y}/\mathcal{E}$ in §3.2) showed a U-shape vs time on the long run – or an inverse U-shape for energy intensity of GDP ($= 1/A = \mathcal{E}/\mathcal{Y}$) seen e.g. p. 350 of ref. 43. The inclusion of biofuels, fodder and food effectively levels the first part of the historical curve as seen at a global level in Fig. A1. A similar feature was also observed for energy intensity in the USA[45], where the hitherto much commented increase of energy intensity disappears and reverses into a steady decrease when traditional energies are included. Whereas some individual countries still exhibit inverse U-shape for time series of energy intensity, it seems that this is a geographical bias: energy intensity never notably increases when consumption is averaged out on large enough areas (energy productivity never decreases). In the European case[46], a similar but partial leveling of energy intensity was obtained by integrating international trade, i.e. subtracting from the UK and German account the energy spent for heavy industrial products sent abroad.

Let us expand on the methodology of traditional fuel estimates. Whereas usage of modern sources of energy (4 to 8) has been accounted for and can be known at the level of individual countries, the consumption of those traditional sources of energy (1 to 3) has been estimated/assessed by Malanima[25] on a macro-area basis (8 geographical macro-areas: Western Europe, Eastern Europe, etc.) as per capita averages. Apart from well tabulated modern energy consumption in the last 100 years, most data are assessed on a decennial basis, or every 5 years[42], and later interpolated on a yearly basis. For details, we refer the reader to Malanima's work, which encompasses/compiles many bibliographical, industrial (e.g. BP) and administrative (e.g. FAO) sources. Malanima also discusses differences with previous long-run estimates especially for fuelwood and compares his aggregate data with other available series. Victor Court also published a complete historical series covering 1800-2014 and the same categories of energy sources, but his work[30] is not cited in Malanima's[25]. Figure A2 compares both aggregate data sets.

Figure A2 shows that data from ref. 25 and 30 are very similar for the whole 20th and 21st centuries, and agree with other sources e.g. BP[10]. However, a noticeable relative gap shows up in the 19th century in Fig. A2. Separate comparisons of the eight energy types reveals that differences between the two sets come essentially from fuelwood. Figure A3 compares specifically those two energy resources.

As previously mentioned, Malanima[25], who does not cite Court[30], compared his fuelwood data with Smil's[31] and noticed that these were much higher than his in the 19th century. The trouble with Court's or Smil's data for biofuels energy consumption – very similar, as shown in Fig. A3a – lies in the per capita numbers, shown in Fig. A4. We remark that Smil's or Court's data imply a decrease of per capita biofuel consumption from 1820 to 1880, at odds with what was happening in the USA, and in numerous other colonial regimes. Moreover, in per capita figures, Court's average global biofuel energy consumption appears higher than the French total PEC per capita, which is very improbable, France being one of the dominant power in the beginning of the period and at the same time consuming mainly wood for heating and industry.

As a whole, like Malanima, we conclude that Smil or Court estimated excessive energy quantities for traditional sources. Concerning food, however, there is a very good agreement between Malanima and Court. For this reason, we retain the data of ref. 25. We can take into account the uncertainty of fuelwood energy figures by using intervals instead of isolated numbers: the curve of Fig. 1, reproduced with error margins in Fig. A5 associated to a 10-50% humidity range, can be seen to shift only gently from this assumption, so such adjustments do not jeopardize our main messages.

*A.2 Global GDP*

Measurement of Gross National Products, then Gross Domestic Products, began during the 1930s in the USA, and generalized progressively to other nations after WWII. Nowadays several international organization (the World bank, OECD, …) provide data for almost every country, which are the basis of a global world product[47].

Gross domestic product (GDP) for a geographical entity using only one currency (e.g. one country) during a short period (one year) is nearly merely a matter of accounting. "GDP is a monetary measure of the market value of all the final goods and services produced in a specific time period"[48]. However, when a complex ensemble of countries, using many different currencies, is surveyed on the long run, the accounting unit need to be corrected[49] for fluctuating exchange rates, inflation and differences in local purchasing power of specific moneys. After these diverse recalibrations of the measurement scale, GDP data are then said to be expressed in a constant currency, usually US$ of a given year, at parity of purchasing power.

For period predating the existence of national accounting, historians try to determine first the income of an average individual in the said country and time. Per capita GDP data are then combined with population data (taken from population census, or from estimates for the periods when no population census was available) to retrieve national GDP, and then global GDP. Within about 1%, there is a consensus over world population estimates from the beginning of the 19th century. We have been using Malanima's population series[25].

Angus Maddison[28] published in 2008 a first estimate of the world product, or global GDP, at selected years since year since the beginning of the Common Era, and for our need: 1820, 1870, 1900, 1913, 1940 and every year since 1950. After his passing away in 2010, his colleagues and followers at University of Groningen continued and improved the work in the so-called "Maddison Project"[50]. New global estimates of GDP were recently released by the team[51,52]. The database provides GDP per capita for nearly every country in US$ 2011 at Parity of Purchasing Power (US$2011 PPP) at selected years. After 1820, data begin to be denser in the (year, country) table. Two different GDP per capita estimates are given: RGDPNA, preferred by the authors for growth measurements, and CGDP, preferred to compare the level of development in different countries. Another database provides aggregate figures for continental regions and the entire world economy[53], shown in Fig. A6.

To study the relations between energy consumption and GDP, yearly data are requested, or at least data at a regular decennial or half-decennial frequency. Thus, several scholars used the Maddison Project databases by year and country to establish yearly global GDP data, from 1870[54], or 1820[25]. Gapminder[55], a well-established statistician enterprise founded by the late Hans Rosling, tackled the task of completing Maddison project database using also other sources: they recently released a new set of data[56]. The new global GDP per capita estimate is labeled gdppc_cppp. Multiplied by the world population, it provides a series of world GDP, shown in Fig. A6. Figure A6 compiles the available world GDP data, showing a convergence of numbers for the part after 1950, and very similar figures for the rest of the studied period. Minute differences observed between series in Fig. A6 would not change the appearance of Fig. 1, and the conclusion of the paper. We retained Gapminder series since it is the only one available on a yearly basis.

**Fig. A1.**

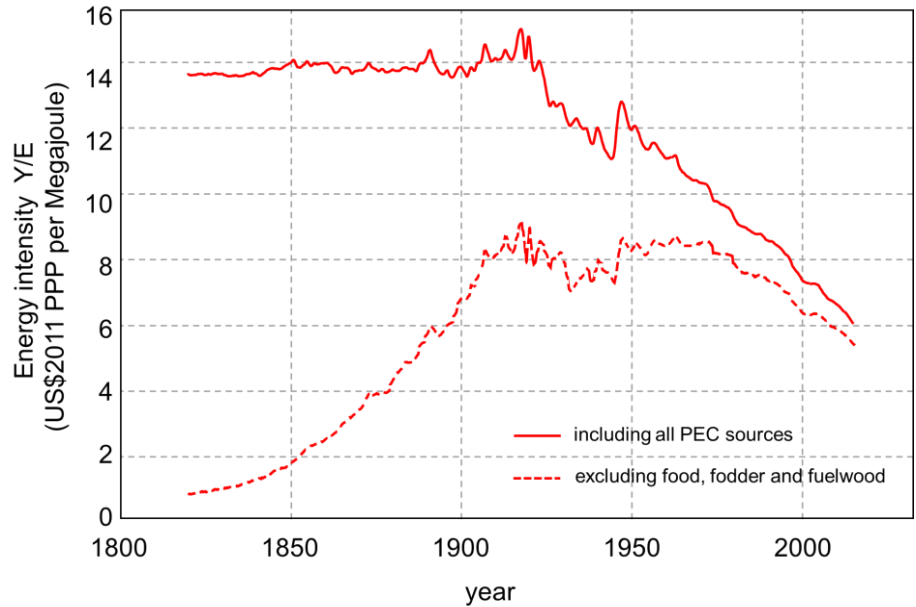

**Fig. A1**: Effect of the inclusion of traditional energy sources (items 1 to 3, food, fuelwood and fodder) on Energy Intensity – i.e. the ratio of primary energy consumption (PEC) to GDP; same data as Fig. 1. Without traditional energies, the curve shows an inverse U-shape. This shape, found very often at the national level, especially when energy sources are only partially accounted for, was the subject of many comments and several theories. When traditional energies are included, plain line, the left part of the curve is levelled.

**Fig. A2.**

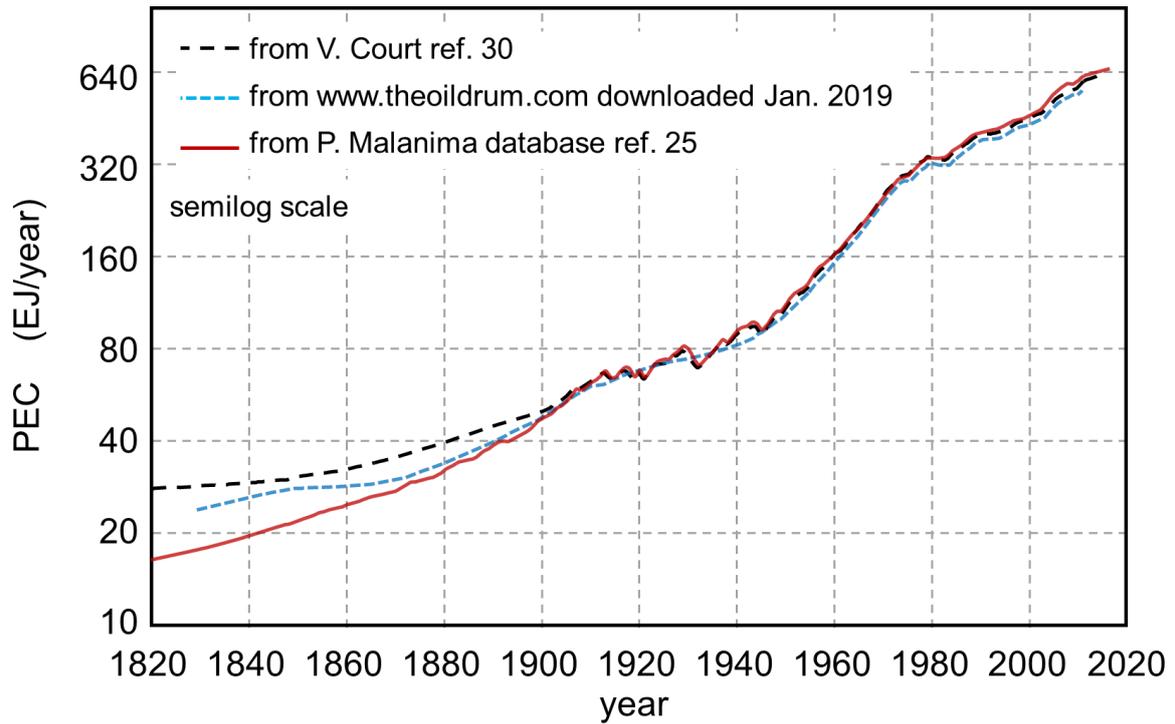

**Fig. A2**: Global Aggregate Energy Consumption from different database - EJ/year. Comparison of Paolo Malanima's database[25], red line, with Victor Court's data[30], dashed black line, and another dataset available on the web, dotted blue line. Energy in EJ/year, semi-log scale. Court's and Malanima's data are very similar in the 20th and 21st centuries. Differences in the 19th century come mainly from fuelwood. Cf. §A.1 and Fig. A3 and A4.

**Fig. A3.**

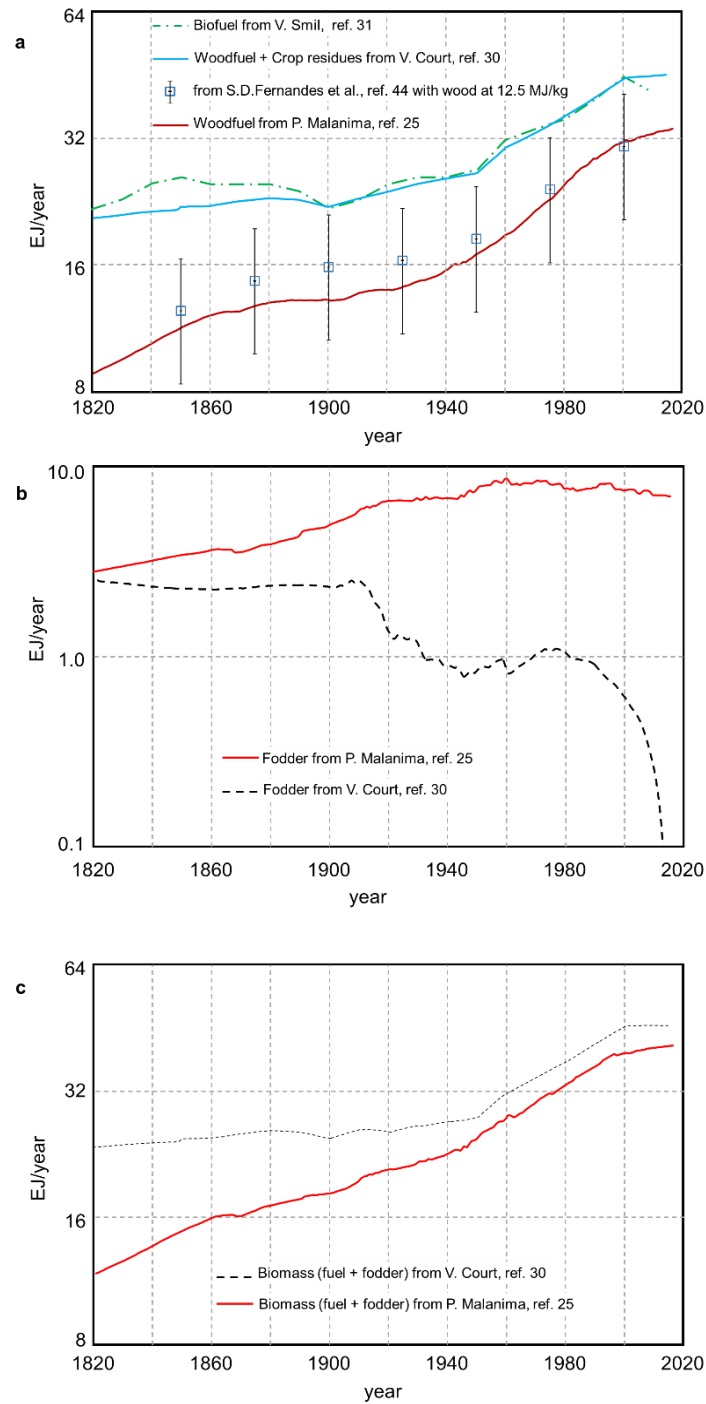

**Fig. A3**: comparison of traditional biomass energy data from Malanima[25] and Court[30]. **a**: solid biofuels series, also compared to Smil's estimate[31] (which only considered solid fuels, not fodder) and to Fernandes's[44] mass of biofuels converted with 12.5 MJ/kg (and a proportional error bar of +/- 20%). **b**: fodder series. **c**: sum of solid fuels and fodder. Court's data are very similar to Smil's for solid fuels (**a**), much higher than Malanima's, whereas the gap between Court's and Malanima's series is reversed for fodder (**b**).

24/09/2020**Fig. A4.**

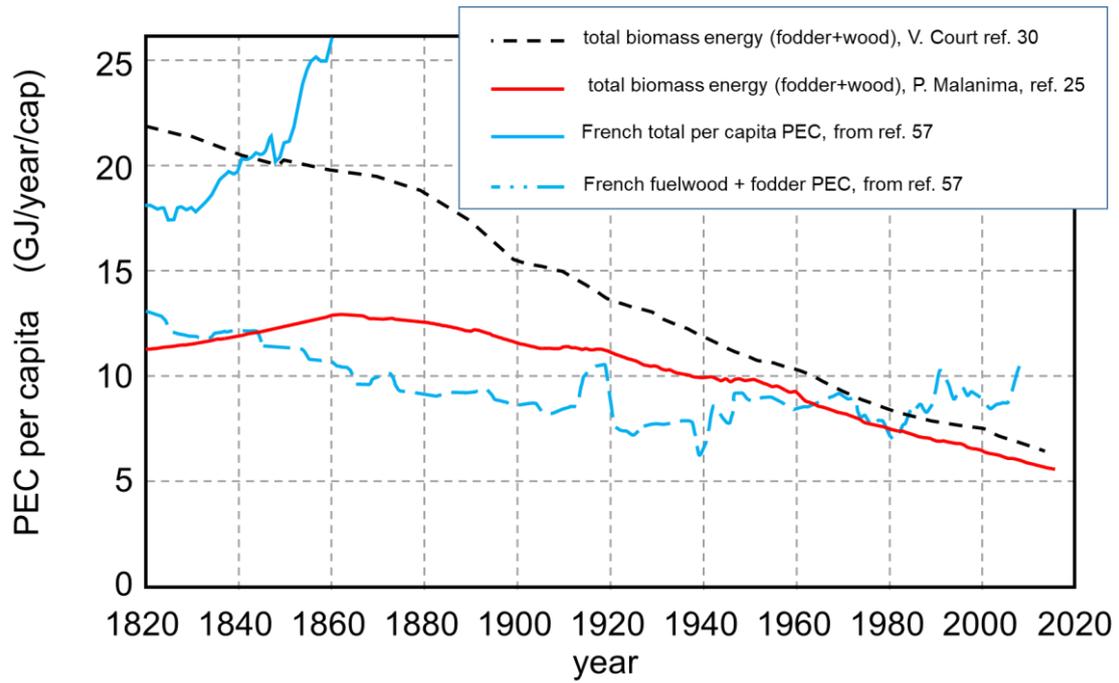

**Fig. A4:** traditional biomass energy per capita. Global data from ref. 25 and 30. French data from the Energy History Project[57]. Court's average global bioenergy consumption per capita is higher than French consumption in the early 19th century, which appears very improbable.

**Fig. A5.**

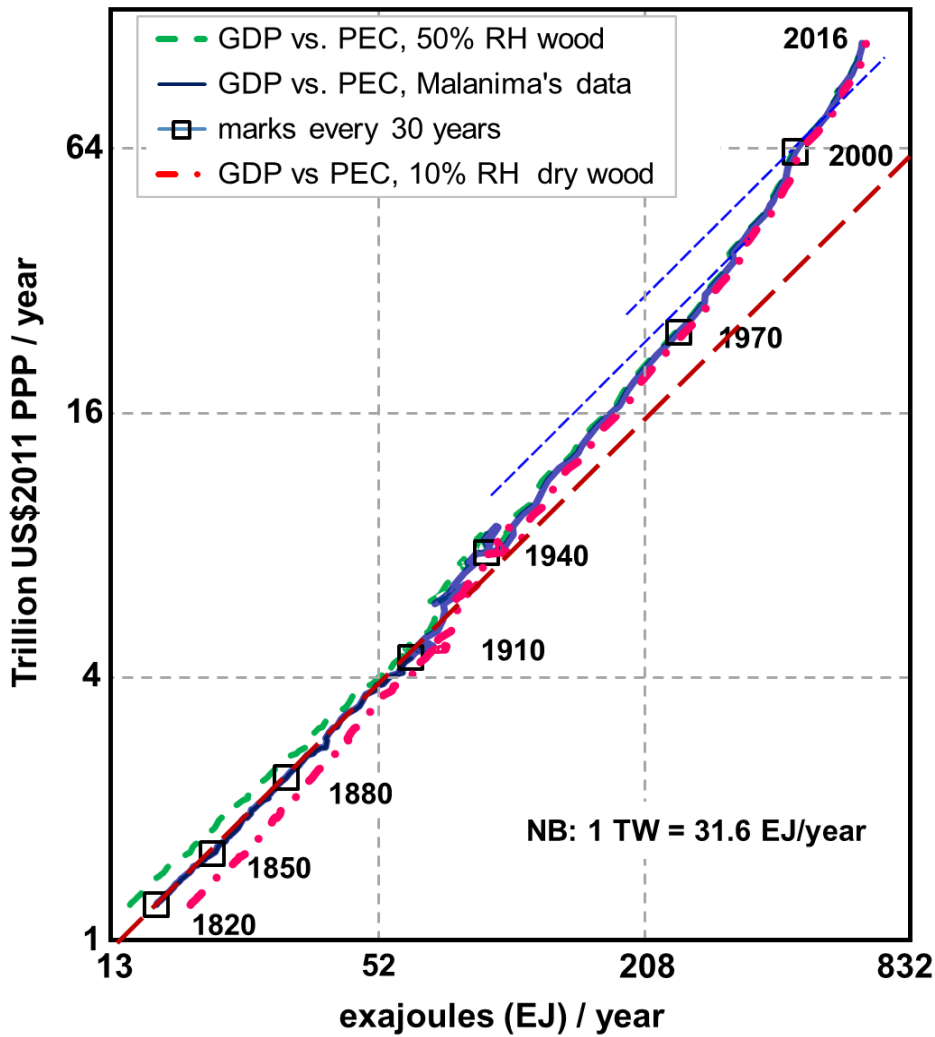

**Fig. A5:** reproduces Fig. 1, allowing for uncertainty in fuelwood energy. This uncertainty is expressed as variability in the average relative humidity [RH] retained to convert Fernandes *et al.*[44] biomass data into energy. Using about 33% RH in wood, which translates into 12.5 MJ/kg wood, Malanima established the aggregate energy consumption series that has been used in the main document. GDP (see §A.2 for a discussion of the data) is plotted vs PEC. The green (respectively red) dashed (resp. point-dashed) line is the same GDP vs aggregate energy using 50% RH, i.e. 9.5 MJ/kg (resp. 10% RH, i.e. 17 MJ/kg). The relative horizontal gap between green and red lines is about ±20% around the Malanima's series. The gap shrinks with time, and growth of modern energy consumption

**Fig. A6.**

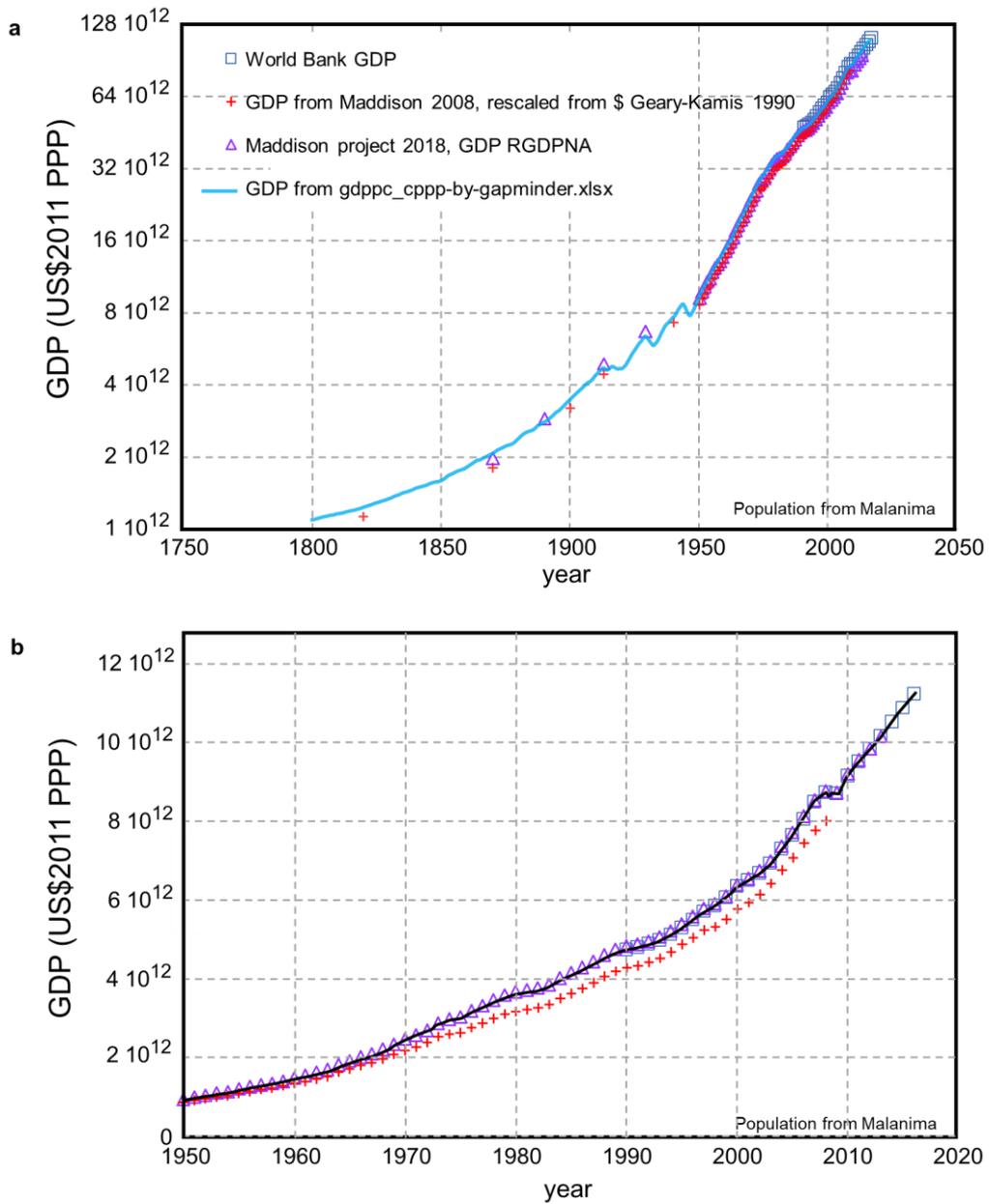

**Fig. A6:** time series of world GDP, from different sources. A6a: semi logarithmic scale, from 1800 to 2016 ; A6b: linear scale, 1950-2016. All sources, except for Maddison 2008, use US$2001 PPP as unit. For Maddison 2008, we rescaled the original data expressed in US$ 1990 dollars PPP to US$2011, using a 1.57 factor correcting inflation, obtained from US inflation data[47]. Apart from a proportional gap of about 10% – which may be due to an incorrect rescaling from our side – between Maddison 2008 and the three other sets of numbers, a general agreement is found. Note that world bank data begin in 1990.

## B. modeling

In appendix B, we go into more details about the resolution of differential equations employed in the main document. In the balanced growth case ($A=1$) corresponding to epoch 1, we solve exactly the system of differential equations for a single capital factor $\mathcal{J}$ in addition to PEC in the Cobb-Douglas function; we briefly discuss the case of two capital factors $H$ and $K$. When $A$ is variable, as in the case of epoch 2, the simplified form used in §3.2 (only energy was kept) is justified here by solving numerically a system with a single capital $\mathcal{J}$ and PEC.

### B.1 balanced growth

Here we give the basic elements to perform the macroeconomic modeling presented in the main document. We follow the spirit and basic mathematics of macroeconomic textbooks[12-14]. Our first attempt used Labor $L$ and capital $K$ as production factors – as usually introduced in textbooks – together with PEC $E$: the production function is then a function of these three factors:

$$Y \propto K^\alpha L^\beta E^\gamma \qquad (B.1)$$

However, the linearity between $Y$ and $E$ (such as in epoch 1), considered together with the investment equation $\frac{dK}{dt} = s_K Y - \delta_K K$ (already seen in Eq. (2) [in appendix B, equations are marked with a B, as above for Eq. (B.1)]) ends up giving a minute role to $L$, in accordance with many studies in the past[16-18]. The disappearance of workers from the equation is counterintuitive and suggests an awkward relation to reality and entailed policies. Two steps heal this problem, as explained in the article. First of all, every human contribution is divided on the one hand into its mere brute force, unskilled labor which only depends on the energy converted by muscles, and on the other hand into a cultural part, prone to improve due to education and knowledge growth. Mankiw et al[32] provided such a theory, where a human capital $H$ appears, at par with the physical, usual capital $K$: it is thus not a disappearance of Labor but a skill-based redefinition. A second step consists in integrating unskilled labor in the aggregate PEC. The combination of both steps produces Eq. (1), repeated here:

$$\frac{Y}{Y_{1820}} = A \left(\frac{K}{K_{1820}}\right)^\alpha \left(\frac{H}{H_{1820}}\right)^\beta \left(\frac{E}{E_{1820}}\right)^\gamma \qquad (1)$$

with $\alpha + \beta + \gamma = 1$. Equation (1) is combined with Eq. (2)

$$\frac{dK}{dt} = s_K Y - \delta_K K \ \& \ \frac{dH}{dt} = s_H Y - \delta_H H \qquad (2)$$

where $Y$, $K$ and $H$ need to be normalized. Using $\mathcal{Y} = \frac{Y}{Y_{1820}}$, $\mathcal{K} = \frac{K}{K_{1820}}$, $\mathcal{H} = \frac{H}{H_{1820}}$, $\mathcal{E} = \frac{E}{E_{1820}}$

it is straightforward to obtain the system:

$$\mathcal{Y} = A\mathcal{K}^\alpha \mathcal{H}^\beta \mathcal{E}^\gamma \qquad (B.2a)$$

$$\frac{d\mathcal{K}}{dt} = \frac{s_K}{\beta_K}\mathcal{Y} - \delta_K \mathcal{K} \qquad (B.2b)$$

$$\frac{d\mathcal{H}}{dt} = \frac{s_H}{\beta_H}\mathcal{Y} - \delta_H \mathcal{H} \qquad (B.2c)$$

with $\beta_K = \frac{K_{1820}}{Y_{1820}}$ and $\beta_H = \frac{H_{1820}}{Y_{1820}}$ the physical and human capital to production ratios at the beginning of the studied period. In the following, we will write $\sigma_K = \frac{s_K}{\beta_K}$ and $\sigma_H = \frac{s_H}{\beta_H}$. With PEC growing exponentially as $\exp[gt]$, it follows logically from the investment equations (2) that $\mathcal{K}$ and $\mathcal{H}$ will grow when energy is consumed. In macroeconomic models with labor $L$, it is usual to set exogenously an exponential growth of $L$, approximated by the global population. In our views, since energy is considered as the primary input, it suffices to set PEC growth. If we define an exponential growth of energy *per capita*, as $\exp[g_\varepsilon t]$, it then implies that population $L$ grows as $\exp[(g - g_\varepsilon)t]$. As usual in macroeconomic textbooks (e.g. ref. 12-14), it is practical to study ratios of production and capital(s) to PEC that are noted $y_\varepsilon = \frac{\mathcal{Y}}{\mathcal{E}}$, $k_\varepsilon = \frac{\mathcal{K}}{\mathcal{E}}$, $h_\varepsilon = \frac{\mathcal{H}}{\mathcal{E}}$. Eq. (B.2a) becomes, still with $\alpha + \beta + \gamma = 1$.

$$y_\varepsilon = A k_\varepsilon^\alpha h_\varepsilon^\beta \qquad (B.3a)$$

Eq. (B.2b) can be rewritten

$$\frac{dk_\varepsilon}{dt} = \frac{d\mathcal{K}}{\mathcal{E}\,dt} - \frac{\mathcal{K}}{\mathcal{E}^2}\frac{d\mathcal{E}}{dt} = \frac{\sigma_K \mathcal{Y} - \delta_k \mathcal{K}}{\mathcal{E}} - \frac{\mathcal{K}}{\mathcal{E}}\frac{d\mathcal{E}}{\mathcal{E}\,dt}$$

which simplifies into

$$\frac{dk_\varepsilon}{dt} = \sigma_K y_\varepsilon - (\delta_K + g)k_\varepsilon \qquad (B.3b)$$

and the same holds for Eq. (B.2c)

$$\frac{dh_\varepsilon}{dt} = \sigma_H y_\varepsilon - (\delta_H + g)h_\varepsilon \qquad (B.3c)$$

For a constant $A = 1$, and for constant values of parameters (saving and depreciation rates), the system of three equations (B.3) can be solved, at least numerically. The resolution of (B.3) with the two types of capital is beyond the scope of this appendix. The simpler case with a unique capital factor is used in §3.1 for a simpler discussion of the dynamics of capital formation, and solved exactly in the following to explicit the steady-state *balanced growth* conditions.

We present here the basics of a simplified system with only one capital noted $\mathcal{J}$ in normalized form. $\mathcal{J}$ and $\mathcal{Y}$ are linked by Eqs. (3) and (4), reproduced here:

$$\mathcal{Y} = \mathcal{J}^\alpha \, \mathcal{E}^{1-\alpha} \tag{3}$$

$$\frac{d\mathcal{J}}{dt} = s\mathcal{Y} - \delta \mathcal{J} \tag{4}$$

For constant values of $s$ and $\delta$, this system is exactly solvable; most textbooks present a discussion of the fixed point only, but an exact solution of the dynamics can also be found elsewhere[13]. A solution can be reached rather easily using production to energy ratio and capital to energy ratio, $y_\varepsilon = \frac{\mathcal{Y}}{\mathcal{E}}$ and $j_\varepsilon = \frac{\mathcal{J}}{\mathcal{E}}$, with which the system is written:

$$y_\varepsilon = j_\varepsilon^\alpha \tag{B.4a}$$

$$\frac{dj_\varepsilon}{dt} = s y_\varepsilon - (\delta + g) j_\varepsilon \tag{B.4b}$$

Combining both equations, we derive a differential equation on $j_\varepsilon$ alone,

$$\frac{dj_\varepsilon}{dt} = s j_\varepsilon^\alpha - (\delta + g) j_\varepsilon \tag{B.5}$$

A few trivial lines bring:

$$\frac{1}{1-\alpha} \frac{d\left(j_\varepsilon^{1-\alpha}\right)}{dt} + (\delta + g) j_\varepsilon^{1-\alpha} = s \tag{B.6}$$

This first order equation on $j_\varepsilon^{1-\alpha}$ is simplified by definition of a new variable $\varphi[t] = j_\varepsilon^{1-\alpha} e^{(1-\alpha)(\delta+g)t}$, which obeys a trivial equation:

$$\frac{d\varphi}{dt} = (1-\alpha) s . e^{(1-\alpha)(\delta+g)t} \tag{B.7}$$

easily integrated as

$$\varphi[t] = 1 + \frac{s}{\delta+g}\left(e^{(1-\alpha)(\delta+g)t} - 1\right) \tag{B.8}$$

since the initial condition (in 1820) is $\varphi[0] = 1$. It is now straightforward to express $j_\varepsilon$

$$j_\varepsilon[t] = \left( e^{-(1-\alpha)(\delta+g)t} + \frac{s}{\delta+g}\left(1-e^{-(1-\alpha)(\delta+g)t}\right) \right)^{\frac{1}{1-\alpha}} \quad (B.9)$$

Capital to energy ratio tends exponentially to a constant $j_\varepsilon^* = \left(\frac{s}{\delta+g}\right)^{\frac{1}{1-\alpha}}$ with characteristic time $\frac{1}{\delta+g}$. From $j_\varepsilon$ and (B.4a), one easily obtains the production to energy ratio:

$$y_\varepsilon[t] = \left( e^{-(1-\alpha)(\delta+g)t} + \frac{s}{\delta+g}\left(1-e^{-(1-\alpha)(\delta+g)t}\right) \right)^{\frac{\alpha}{1-\alpha}} \quad (B.10)$$

which converges with a characteristic time $\frac{1}{\alpha(\delta+g)}$ to

$$y_\varepsilon^* = \left(\frac{s}{\delta+g}\right)^{\frac{\alpha}{1-\alpha}} \quad (B.11)$$

In the asymptotic regime, called "balanced growth" in macroeconomic textbooks[12-14], capital and production are proportional to PEC and grow exponentially according to $\exp[gt]$.

Their asymptotic ratio is $\frac{J}{Y} = \frac{j_\varepsilon^*}{y_\varepsilon^*} = \frac{\left(\frac{s}{\delta+g}\right)^{\frac{1}{1-\alpha}}}{\left(\frac{s}{\delta+g}\right)^{\frac{\alpha}{1-\alpha}}} = \frac{s}{\delta+g}$ which is fixed when the parameters $\delta$, $s$ and $g$ are constant. Variation of these parameters, on the other hand, would produce an evolution of the capital to production ratio: its mathematical study calls for a new, more complex resolution of (B.4a) and (B.4b), which will not be tackled here. We will only give a brief discussion of epoch 1, extending the resolution of (B.4) to the case of the system (B.3) of three equations.

Epoch 1 is characterized by such a balanced growth dynamics. Similarly to system (B.4), (B.3b) and (B.3c) imply the convergence of $k_\varepsilon$ to a fixed point $k_\varepsilon^* = \left(\frac{\sigma_K}{g+\delta_K}\right)^{\frac{1-\beta}{1-\alpha-\beta}} \left(\frac{\sigma_H}{g+\delta_H}\right)^{\frac{\beta}{1-\alpha-\beta}}$, while $h_\varepsilon$ converges to $h_\varepsilon^* = \left(\frac{\sigma_H}{g+\delta_H}\right)^{\frac{1-\alpha}{1-\alpha-\beta}} \left(\frac{\sigma_K}{g+\delta_K}\right)^{\frac{\alpha}{1-\alpha-\beta}}$. Compared to the simpler case with one capital, the dynamics will be made richer due to several different characteristic times of convergence. Since empirical data imply that $y_\varepsilon$ is constant, $y_\varepsilon = \frac{Y}{Y_{1820}} \frac{E_{1820}}{E} = 1$, the actual asymptotic relations can be expressed, without much surprise, by simpler relations, namely: $k_\varepsilon^* = \dfrac{\sigma_K}{g+\delta_K}$ and $h_\varepsilon^* = \dfrac{\sigma_H}{g+\delta_H}$. Reintroducing the non-normalized $K$ and $H$, as well as $Y$ and $E$, remembering $\beta_K = \dfrac{K_{1820}}{Y_{1820}}$ and $\sigma_K = \dfrac{s_K}{\beta_K}$, we retrieve $\dfrac{K}{Y} = \dfrac{s_K}{\delta_K + g}$ the fixed-point ratio of physical capital to production, which goes also here with the same relation for the human capital $\dfrac{H}{Y} = \dfrac{s_H}{\delta_H + g}$.

### B.2 case of a variable $A$ in epoch 2

In epoch 2, there is no more linearity of GDP with PEC. This can be modeled with a variable $A$ in Eq. (1). In §3.2, this situation is studied in a very simplified version where the capital factors $K$ and $H$ are put aside, and where $E$ only is kept: see Eq. (5) to (7). Here we consider a compromise solution, extending the single capital model (Eq. (3) and (4)) to the case of a variable residual $A$.

$$\mathcal{Y} = A \mathcal{J}^\alpha \mathcal{E}^{1-\alpha} \qquad (B.12a)$$

$$\dfrac{d\mathcal{J}}{dt} = s\mathcal{Y} - \delta \mathcal{J} \qquad (B.12b)$$

where we used normalized quantities $\mathcal{Y} = \dfrac{Y[t]}{Y[t_0]}$, $\mathcal{J} = \dfrac{J[t]}{J[t_0]}$, etc. The simplified case treated in §3.2 can be retrieved by taking $\alpha = 0$. As previously, we anticipate that $A$ is related to PEC by

$$\dfrac{d\text{Log}[A]}{dt} = \chi_\alpha \mathcal{E} \qquad (B.12c)$$

with a constant $\chi_\alpha$. We have $\chi_{\alpha=0}$, the value of $\chi_\alpha$ for $\alpha = 0$, that corresponds to $\chi$ obtained above in §3.2. We take $\mathcal{E}$ as exogenously defined and growing exponentially according to $\exp[g(t-t_0)]$ from a time $t_0 = 0$. (B.12c) is directly transformed into $A[t] = \exp[\int_{t_0}^t \chi_\alpha \mathcal{E}\, dt] = \exp[\int_0^t \chi_\alpha \exp[gt]\, dt]$. To solve numerically the system (B.12), we go through the same process as above: we study $y_\varepsilon = \dfrac{\mathcal{Y}}{\mathcal{E}}$ and $j_\varepsilon = \dfrac{\mathcal{J}}{\mathcal{E}}$ which obey the system of equations (B.13)

$$y_\varepsilon = \exp[\dfrac{\chi_\alpha}{g}\exp[gt]] j_\varepsilon^\alpha \qquad (B.13a)$$

$$\frac{dj_\varepsilon}{dt} = sy_\varepsilon - (\delta + g) j_\varepsilon \qquad (B.13b)$$

At the beginning of the period, at time $t_0$, we suppose that the conditions of balanced growth are met: this implies that $\frac{j_\varepsilon[t_0]}{y_\varepsilon[t_0]} = \frac{s}{\delta + g}$. But we also have $y_\varepsilon[t_0] = j_\varepsilon[t_0] = 1$. Thus, we have $s = \delta + g$. In a last simplifying step, we define the temporal scale such that s equals 1.

Figure B1 shows the evolution of $y_\varepsilon = \frac{\mathcal{Y}}{\mathcal{E}}$ $j_\varepsilon = \frac{\mathcal{J}}{\mathcal{E}}$ and $A$ for $\alpha = 0.5$. On B, we see that all quantities scale like an exponential function of $\int_{t_0}^{t} \mathcal{E} dt$ within less than a percent. The residual $A$ is such through our initial hypothesis ($A[t] = \exp[\int_{t_0}^{t} \chi_\alpha \mathcal{E} dt]$). Figure B1b proves for $\alpha = 0.5$ that this hypothesis results in the expected behavior for $y_\varepsilon = \frac{\mathcal{Y}}{\mathcal{E}}$ which was found to scale like $\exp[\int_{t_0}^{t} \mathcal{E} dt]$ (see §3.2, especially Fig. 2). The parameter $\chi_{0.5}$ in Fig. 7 was chosen to provide a factor 2.3 increase of $y_\varepsilon = \frac{\mathcal{Y}}{\mathcal{E}}$ as in the case of the historical period 1820-2016. The behavior of $y_\varepsilon$ when $\alpha$ varies is very stable. Its limited range can be seen for example in Fig. B2a where the same functions as Fig. B1b are plotted, here for $\alpha = 1$ with $\chi_1 = 0.0011$. In Fig. B2b, we plot in semilog scale the production to energy ratio $y_\varepsilon = \frac{\mathcal{Y}}{\mathcal{E}}$ for the two values of $\alpha$: a minute gap, smaller than 3%, is found between the two curves.

We performed simulations for other values of $\alpha$ in the range 0 to 1, with very minor changes in the results. This shows that the hypothesis $A[t] = \exp[\int_{t_0}^{t} \chi_\alpha \mathcal{E} dt]$ with a constant $\chi_\alpha$ correctly renders the historical period 1820-2016, whatever the value of $\alpha$ in Eq. (B.12a).

If one would reproduce the final discussion of §3.2 in the more general case of a model including a capital factor $\mathcal{J}$, similar conclusions would be reached.

**Fig. B1.**

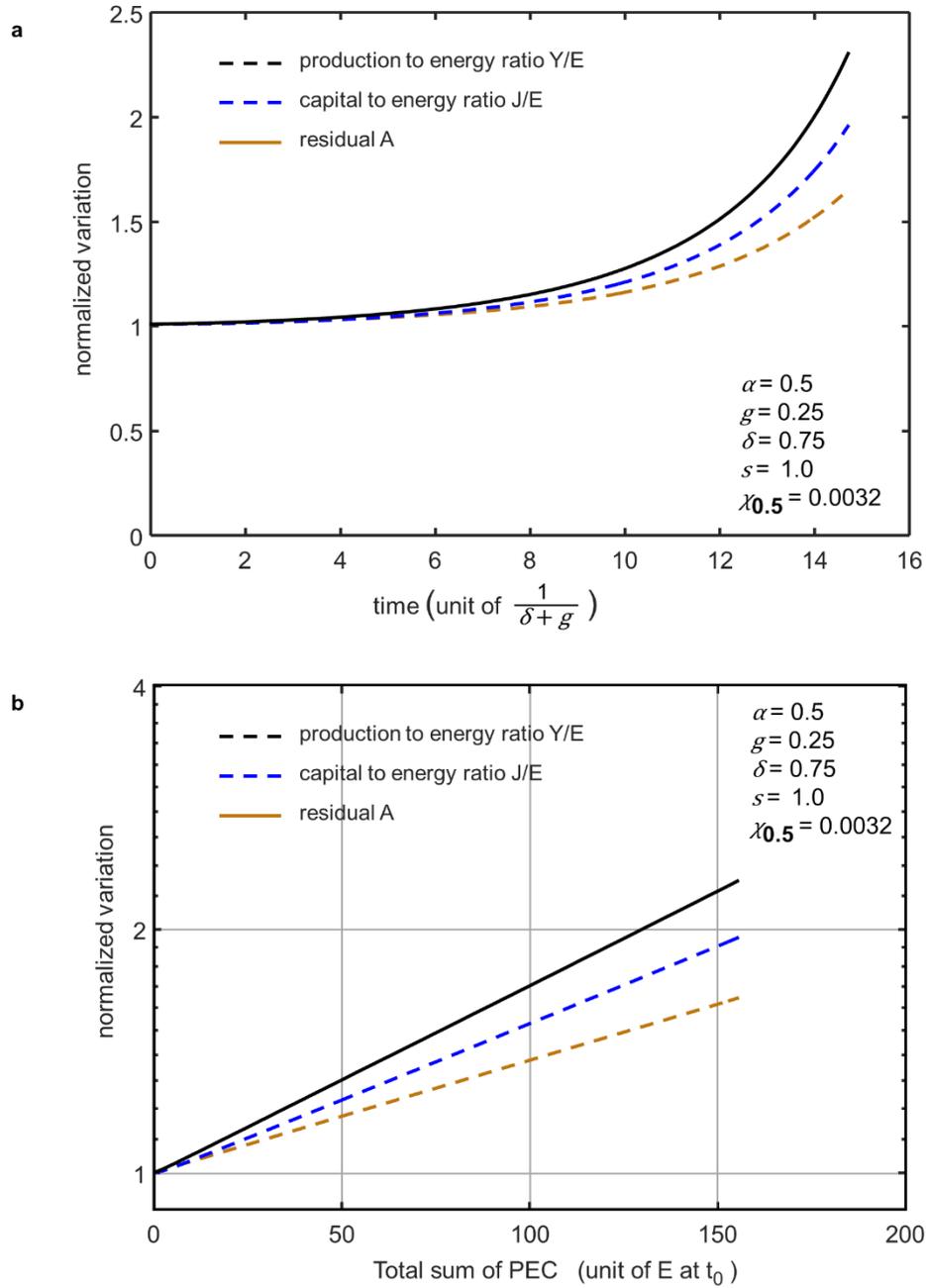

**Fig. B1:** Production to energy ratio $y_\varepsilon = \frac{Y}{\mathcal{E}}$, the same ratio for its capital factor $j_\varepsilon = \frac{J}{\mathcal{E}}$ and the residual $A$, calculated for $\alpha = 0.5$ according to Eq. (B.13). **a**: plotted on a linear scale as a function of time, normalized to $\frac{1}{\delta+g}$, the effective depreciation time. **b**: plotted on a semilog scale as a function of the total time integral of PEC, normalized to PEC at $t_0 = 0$, with the same time unit as B1a.

**Fig. B2.**

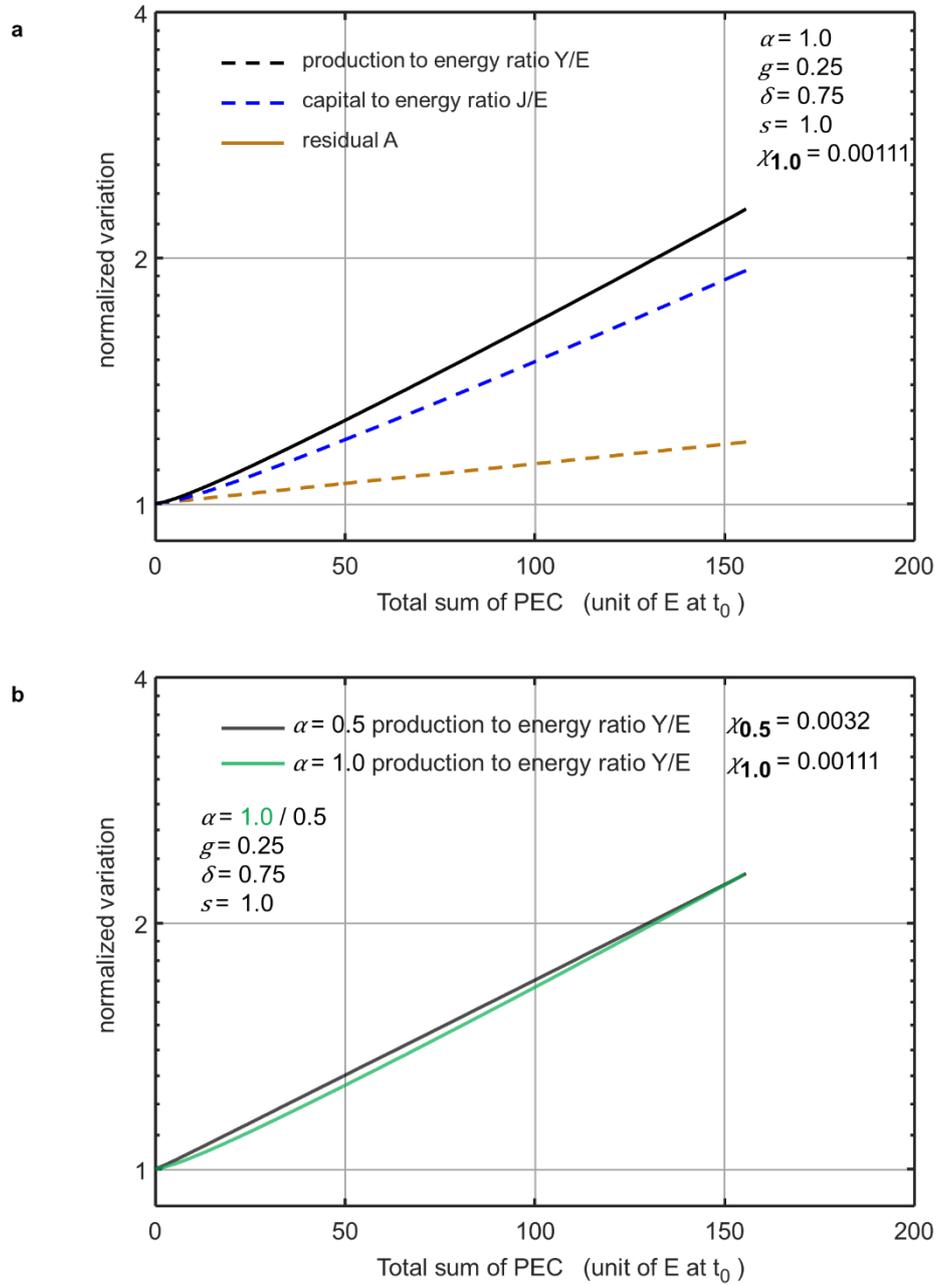

**Fig. B2:** Production to energy ratio $y_\varepsilon = \dfrac{\mathcal{Y}}{\mathcal{E}}$, the same ratio for its capital factor $j_\varepsilon = \dfrac{\mathcal{J}}{\mathcal{E}}$ and the residual $A$, calculated for $\alpha = 1$, according to Eq. (B.13). **a**: plot on a semilog scale as a function of the total sum of PEC, same abscissa unit (energy) as Fig. B1b. **b**: comparison of $y_\varepsilon = \dfrac{\mathcal{Y}}{\mathcal{E}}$ on a semilog scale for the two values of $\alpha$; both curves are close to a straight line.

24/09/2020